# Simultaneous Deep Learning of Myocardium Segmentation and $T_2$ Quantification for Acute Myocardial Infarction MRI

Yirong Zhou [#], Chengyan Wang [#], Mengtian Lu, Kunyuan Guo, Zi Wang, Dan Ruan, Rui Guo, Peijun Zhao, Jianhua Wang, Naiming Wu, Jianzhong Lin, Yinyin Chen, Hang Jin, Lianxin Xie, Lilan Wu, Liuhong Zhu, Jianjun Zhou, Congbo Cai, He Wang, Xiaobo Qu *

*Abstract*—In cardiac Magnetic Resonance Imaging (MRI) analysis, $T_2$ mapping is a tissue quantification technology to measure water and inflammation levels and has been recognized as an important versatile index for myocardial pathologies. The $T_2$ quantification is a highly relevant task with the myocardial segmentation but they only have been explored separately in deep learning. Here, we proposed a simultaneous dual-task network for myocardial segmentation and $T_2$ quantification, called SQNet. It integrates Transformer and Convolutional Neural Network (CNN) components. SQNet has a $T_2$-refine fusion decoder for quantitative analysis, incorporating global features extracted by the Transformer, and employs a segmentation decoder with multiple local region supervision to enhance segmentation accuracy. Additionally, SQNet proposes a tight coupling module that aligns and fuses CNN and Transformer branch features. The mutual promotion of the two tasks enables the SQNet to focus on the myocardium regions. The dual-task performance is validated on healthy controls (HC, N=17) and acute myocardial infarction patients (AMI, N=12). The segmentation dice of SQNet on HC/AMI is 89.3/89.2, which is higher than that (87.7/87.9) of the compared state-of-the-art segmentation method. The Pearson correlation coefficient of the $T_2$ value on HC/AMI is 0.84 and 0.93 to the label value, indicating a strong linear correlation in myocardial $T_2$ mapping. From the clinical diagnostic perspective, eight experienced radiologists have evaluated the image quality (Score standard: Good is 4.0 and excellent is 5.0) of SQNet on HC/AMI with the segmentation score 4.60/4.58 and the $T_2$ quantification score 4.32/4.42, which is higher than the compared state-of-the-art segmentation score (4.50/4.44) and the $T_2$ quantification score (3.59/4.37). Thus, SQNet provides an accurate simultaneous segmentation and quantification, providing the more accurate diagnosis of cardiac disease, such as AMI.

## I. INTRODUCTION

ACUTE myocardial infarction (AMI) is a serious cardiovascular disease [1]. The myocardium is a muscle tissue inside the heart and is responsible for pumping blood to meet the oxygen and nutritional needs of the entire body [2]. Once a coronary artery blockage occurs, the myocardium is rapidly damaged [3], behavioring as myocardial edema [4] and leading to AMI.

To characterize the myocardial tissue, Magnetic Resonance Imaging (MRI) is a primary noninvasive imaging modality [5, 6]. For example, the $T_2$-weighted ($T_2$w) MRI image provides rich tissue contrast to reveal abnormal changes within myocardial tissue and has reliably identified regions of myocardial edema [7-10]. For AMI, accurately segmenting the myocardium region is important for diagnosis assistance [11-15].

Through acquiring a series of $T_2$w MRI images, a quantitive measure, the $T_2$ mapping, can provide the relaxation time constant in MRI. $T_2$ mapping is a valuable tissue quantification to measure water and inflammation levels [16, 17]. It has been recognized as the versatile index for a range of myocardial pathologies, including myocardial edema [18-20]. The series of $T_2$w images are acquired by varying the time of echo (TE) and then the $T_2$ value is computed at each pixel by fitting an exponential function that models the relationship of pixel value on TE. The most typical fitting method is the nonlinear least

This work was supported by the National Natural Science Foundation of China (62331021 and 62122064), the Natural Science Foundation of Fujian Province of China (2023J02005), Industry-University Cooperation Projects of the Ministry of Education of China (231107173160805), National Key Research and Development Program of China (2023YFF0714200), the President Fund of Xiamen University (20720220063), and the Nanqiang Outstanding Talents Program of Xiamen University. ([#] Equal contribution, * Corresponding author. E-mail address: quxiaobo@xmu.edu.cn)

Yirong Zhou, Kunyuan Guo, Zi Wang, Dan Ruan, Congbo Cai, and Xiaobo Qu are with the Department of Electronic Science, Intelligent Medical Imaging R&D Center, Fujian Provincial Key Laboratory of Plasma and Magnetic Resonance, National Institute for Data Science in Health and Medicine, Xiamen University, Xiamen 361104, China.

Chengyan Wang, and He Wang are with the Human Phenome Institute, Fudan University, Shanghai 201210, China.

Mengtian Lu is with the Department of Radiology, Xianning Central Hospital (The First Affiliated Hospital of Hubei University of Science and Technology), Xianning 437100, China.

Rui Guo is with the School of Medical Technology, Beijing Institute of Technology, Beijing 100081, China.

Peijun Zhao, and Jianhua Wang are with the Department of Radiology, The First Affiliated Hospital of Xiamen University, School of Medicine, Xiamen University, Xiamen 361003, China.

Naiming Wu is with the Department of Radiology, Xiamen Cardiovascular Hospital Xiamen University, School of Medicine, Xiamen University, Xiamen 361004, China.

Jianzhong Lin is with the Department of Radiology, Zhongshan Hospital Affiliated to Xiamen University, School of Medicine, Xiamen University, Xiamen 361004, China.

Yinyin Chen, and Hang Jin are with the Department of Radiology, Zhongshan Hospital, Shanghai 200032, China.

Lianxin Xie, Lilan Wu, Liuhong Zhu, and Jianjun Zhou are with the Department of Radiology, Zhongshan Hospital, Fudan University (Xiamen Branch), Fujian Province Key Clinical Specialty Construction Project (Medical Imaging Department), Xiamen Key Laboratory of Clinical Transformation of Imaging Big Data and Artificial Intelligence, Xiamen 361006, China.

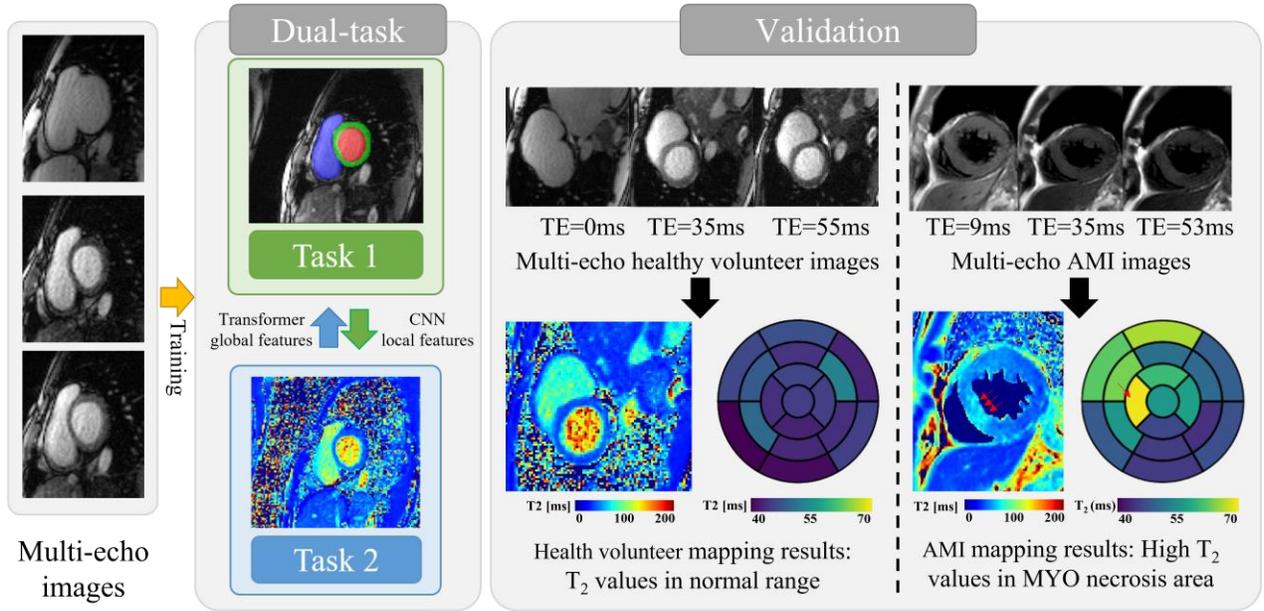

**Fig. 1.** The workflow of dual-task magnetic resonance cardiac segmentation and quantification.

squares (NLLS) [21, 22] and the state-of-the-art one is the deep learning method [23], where the latter can greatly speed up the computation of the former.

However, the deep learning is hard to quantify the $T_2$ value because the single-pixel fitting is easily influenced by the low signal-to-noise-ratio (SNR) or image artifacts [24, 25]. Thus, homogeneous information such as similar $T_2$ values in a local region may enhance the robustness of $T_2$ estimation to the noise or image artifacts.

Although myocardial segmentation [26, 27] and tissue quantification [28] are two highly relevant tasks for AMI, there is no existing research that integrates these two tasks in a single deep learning model. We aim to improve the $T_2$ quantification by inferring the homogeneous information with segmentation.

Here, we proposed an end-to-end simultaneous segmentation and $T_2$ quantification for cardiac MRI, called SQNet. By inputting multi-echo short-axis MRI images, SQNet simultaneously completes segmentation and quantification of the cardiac areas, including the Left Ventricle (LV), Myocardium (MYO), and Right Ventricle (RV). We used a hybrid Transformer-CNN network structure to make segmentation and quantification more accurate than other single-task methods. Besides, a dynamic weight average loss function is adopted to realize the complementary features of dual-task. SQNet will be trained by using data from healthy controls and AMI patients (Fig. 1). The dual-task performance will be validated on the 17 healthy controls and 12 AMI patients and subjectively evaluated by eight radiologists.

## II. PROPOSED METHOD

Here, we described key network components of the SQNet and how to train it.

### A. Network Architecture of SQNet

SQNet consists of two parallel modules that accomplish segmentation and quantification tasks (Fig. 2(a)), respectively.

1) **$T_2$ Quantification Module**

The quantification task adopts an encoder-decoder network to predict the $T_2$ mapping.

In the encoder, we used the transformer block [29]. Each block involved a layer normalization, a multiple-headed self-attention module, and a multilayer perceptron block.

In the decoder, a $T_2$-refine fusion decoder (Fig. 2(b)) was designed, incorporating dropout, three Conv blocks, batch normalization, and a ReLU function. In the last layer, an upsampling block is used to output the $T_2$ mapping result as an image.

To minimize the quantification error, the structural similarity (SSIM) [30-33] between the predicted $T_2$ map $\mathbf{M}$ and the ground-truth $T_2$ map $\mathbf{G}$ is defined as the loss function $\mathcal{L}_{\text{Quant}}$:

$$\mathcal{L}_{\text{Quant}}(\mathbf{G},\mathbf{M}) = 1 - \text{SSIM}(\mathbf{G},\mathbf{M}), \quad (1)$$

where SSIM lies in the range of [0,1]. Here, the ground-truth $T_2$ map is obtained by the traditional NLLS [21, 22]. The computation of NLLS is very long and costs 367.1 seconds to quantify each subject that has 6 slices by running a least_squares function of a python scipy module on a platform with 12 cores of Intel(R) Xeon(R) Silver 4310 CPU.

2) **Segmentation Module**

The segmentation task adopts an encoder-decoder network, using a CNN block encoder and our proposed region supervision decoder (Fig. 2(c)), to predict the segmentation probability maps and then pass this map to the argmax function to obtain the four-class segmentation map.

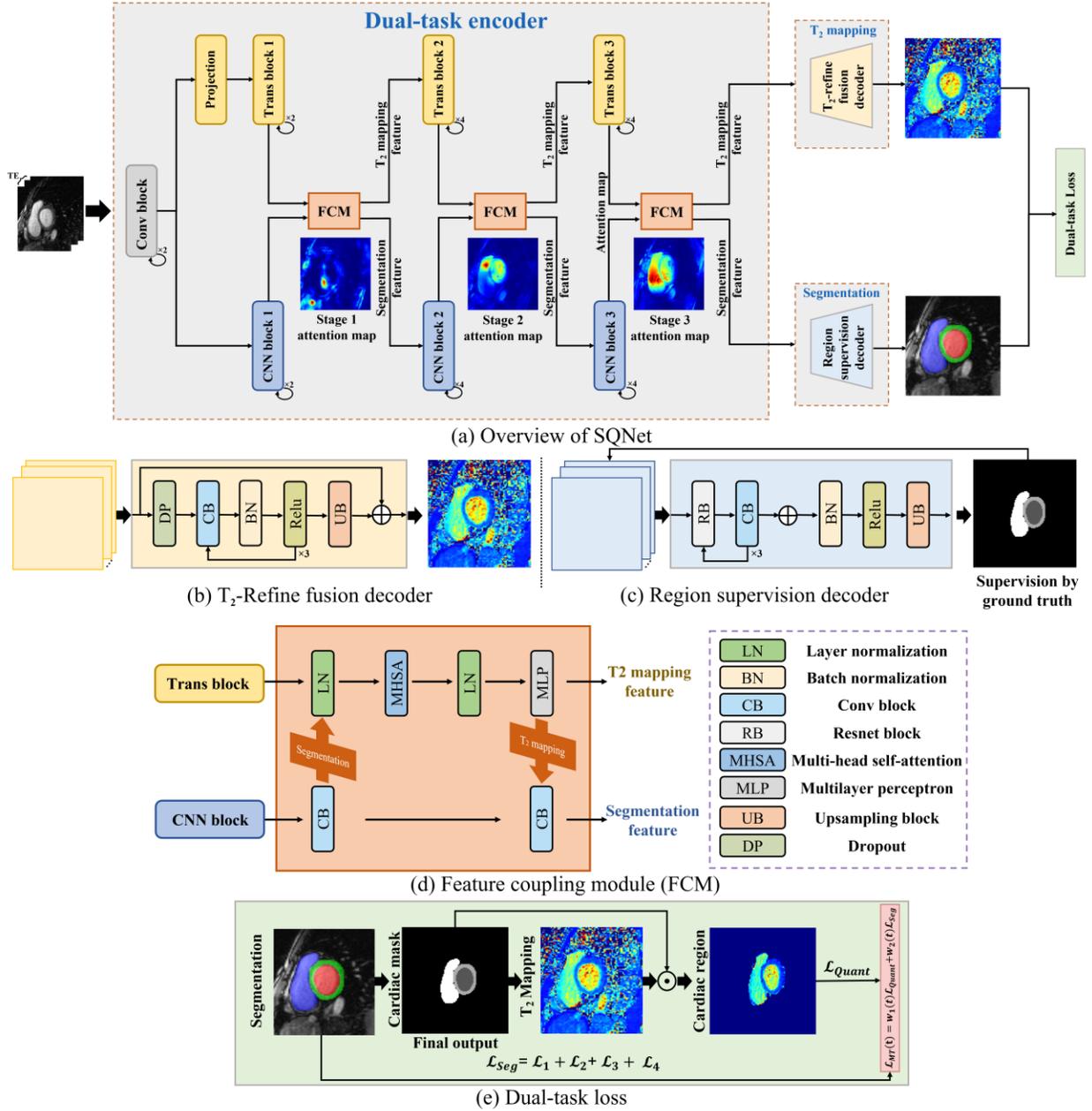

**Fig. 2.** The network architecture of SQNet. FCM denotes feature coupling module.

In the encoder, we used a progressively downsampling CNN blocks architecture, reducing spatial dimensions while increasing channel dimensions. Each block involved multiple bottlenecks of ResNet [34], including down-projection convolution, spatial convolution, and up-projection convolution.

In the decoder, a region supervision decoder was employed (Fig. 2(c)), incorporating three ResNet blocks, three Conv blocks, a batch normalization, and a ReLU function. To enable region supervision (Fig. 3), a 3×3 bilinear interpolation was applied to the last layer of each decoder, aligning with the manual annotations. Therefore, the loss function of the segmentation module contains four different resolutions, and the overall loss is their accumulation.

The segmentation task outputs four probability maps, which correspond to LV, MYO, RV, and background region,

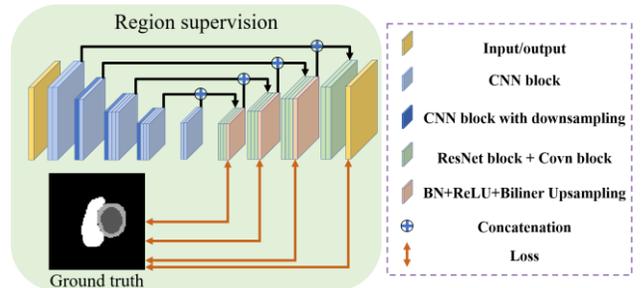

**Fig. 3.** The detail architecture of region supervision.

TABLE I
DATA ACQUISITION PARAMETERS

| Subject | Number | Manufacturer | Sequence name | Acquisition matrix | Slice number | Slice thickness (mm) | Time echo (ms) |
|---|---|---|---|---|---|---|---|
| Healthy control | 120 | 3T Siemens Vida | $T_2$-prepared-Flash | $227 \times 227$ | 5~6 | 5.0 | 0/35/55 |
| AMI patient | 49 | 3T Philips Ingenia | $T_2$-prepared-GRASE | $152 \times 149$ | 3 | 8.0 | 9/28/56 |

Note: Healthy control denotes normal bright blood healthy volunteers, AMI patient denotes black blood acute myocardial infarction.

respectively. The multi-class segmentation loss function $\mathcal{L}_{Seg}$ is defined in the form of cross-entropy:

$$\begin{cases} \mathcal{L}_n(\mathbf{Y},\mathbf{P}) = -\frac{1}{HW}\sum_{h=1}^{H}\sum_{w=1}^{W}\sum_{c=1}^{C}\mathbf{Y}_{nijc}\log(\mathbf{P}_{nijc}) \\ \mathcal{L}_{Seg}(\mathbf{Y},\mathbf{P}) = \frac{1}{N}\sum_{n=1}^{N}\mathcal{L}_n(\mathbf{Y},\mathbf{P}) \end{cases}, \quad (2)$$

where $N$ denotes the different resolution images. $H$ and $W$ denotes the height and width of the images. $C$ denotes the four different classes. $\mathbf{Y}_{nijc}$ and $\mathbf{P}_{nijc}$ denote the label and predict probability belonging to category $c$ at the position $(i, j)$ in the $n^{th}$ image, respectively. The cross-entropy is adopted since it works better for multi-classification problems [35, 36].

### B. Feature Coupling Module

The feature coupling module (FCM) (Fig. 2(d)) combines two key elements: One is the feature map of the segmentation task captured by the local convolution operator, and the other is the feature map of the $T_2$ mapping task captured by the global self-attention mechanism aggregated of the Transformer patch embedding. These two elements are aligned and coupled to enable the network to focus on the heart region, which is shown in Fig. 2(a) three stage attention maps, thereby achieving more accurate segmentation and $T_2$ mapping.

### C. Dual-Task Loss

The dual-task loss function (Fig. 2(e)) is defined as:
$$\mathcal{L}_{MT}(t) = w_1(t)\mathcal{L}_{Quant} + w_2(t)\mathcal{L}_{Seg}, \quad (3)$$

where $t$ denotes the current epoch number, $w_1$ and $w_2$ is a weight to balance the quantification and segmentation loss. The weight is automatically adjusted following a dynamic weight average scheme [37]:

$$\begin{cases} w_1(t) = \dfrac{C_3 \times \exp(w_1(t-1)/C_4)}{\exp\left(\dfrac{\mathcal{L}_{Quant}(t-1)}{\mathcal{L}_{Quant}(t-2)}/C_4\right) + \exp\left(\dfrac{\mathcal{L}_{Seg}(t-1)}{\mathcal{L}_{Seg}(t-2)}/C_4\right)} \\ w_2(t) = \dfrac{C_3 \times \exp(w_2(t-1)/C_4)}{\exp\left(\dfrac{\mathcal{L}_{Quant}(t-1)}{\mathcal{L}_{Quant}(t-2)}/C_4\right) + \exp\left(\dfrac{\mathcal{L}_{Seg}(t-1)}{\mathcal{L}_{Seg}(t-2)}/C_4\right)} \end{cases}, \quad (4)$$

where $\dfrac{\mathcal{L}_{Quant}(t-1)}{\mathcal{L}_{Quant}(t-2)}$ and $\dfrac{\mathcal{L}_{Seg}(t-1)}{\mathcal{L}_{Seg}(t-2)}$ calculate the relative descending rate of different tasks and both of them are initialized as 1. The $C_3 = 1$ and $C_4 = 2$ controls the softness of weighting of two tasks [38].

### D. Dual-Task Network Training

1) **Dataset**

The 2D short-axis cardiac $T_2$ mapping data, including healthy cases from a public CMRxRecon dataset ($N$=120) [39] and an in-house AMI patient ($N$=49) dataset collected by Fudan University. The study protocol was approved by the institutional review board and informed consent was obtained from volunteers before examination. The approving institution is Fudan university and Pudong hospital, the IRB/ethics board protocol number is 2024-MS-R-23, and the date of approval is 2023.07. The training data include partial healthy controls ($N$=103) and AMI patients ($N$=37) and the rest data are used for validation. The data acquisition parameters are listed in Table I.

2) **Data Annotation**

All cardiac images from healthy volunteers were manually annotated with myocardium and chamber labels by a radiologist (with 5 years of cardiac imaging experience) using ITK-SNAP [40] (version 3.8.0).

All cardiac images from AMI patients were manually annotated with myocardium and chamber labels by a radiologist (with 4 years of medical imaging experience) using our CloudBrain-LabelAI system [41] (https://csrc.xmu.edu.cn/CloudBrain.html).

For each $T_2$ map, three $T_2$w images were acquired under three TEs (0 ms, 35 ms, and 55 ms). We chose the $T_2$w image with TE=35 ms to annotate LV, MYO, and RV regions since this image had better contrast than the other $T_2$w images with TE=0 ms and 55 ms. We assumed that the $T_2$w image with a middle-valued TE could represent the other TEs for annotation.

3) **Data Pre- and Post-Processing**

The input cardiac image is denoted as $\mathcal{X} \in \mathbb{R}^{C \times H \times W}$ where $C$, $H$ and $W$ are the number of TEs, the height and the width of cardiac images, respectively. The input image is normalized following the Min-Max rule [42] defined as:

$$\mathcal{X}_{norm} = \frac{\mathcal{X} - \min(\mathcal{X})}{\max(\mathcal{X}) - \min(\mathcal{X})}, \quad (5)$$

where the min and max denote the minimal and maximal pixel intensity of $T_2$w images under multiple TEs. With the cardiac area as the center, center cropping is applied to obtain a square matrix of the input cardiac image [43].

$T_2$ mapping labels $T_2(h,w)$ are computed using the NLLS according to:

$$\min_{T_2(h,w)} \sum_{c=1}^{C} (\mathcal{X}_{norm}^{\{0\}}(h,w) * \exp(-TE_c / T_2(h,w)) - \mathcal{X}_{norm}^{\{c\}}(h,w))^2, \quad (6)$$

where $\mathcal{X}_{norm}^{\{c\}} \in \mathbb{R}^{H \times W}$ is a T2w image under a $TE_c$ of the tensor $\mathcal{X}_{norm}$, where $c \in \{1,2,\cdots,C\}$, $h \in \{1,2,\cdots,H\}$ and $w \in \{1,2,\cdots,W\}$ denotes the TEs index, image height, and width index, respectively. $\mathcal{X}_{norm}^{\{0\}}$ represents the initial magnetization strength, $T_2$ is the $T_2$ map to be found. We performed window truncation to let the $T_2$ value falls within the empirical interval [0 ms, 200 ms].

Data augmentation was performed by random rotation: horizontal/vertical flipping, random cropping, and applying elastic deformation onto the training images. This process increased the number of training samples by 8 times.

4) **Training Details**

We used an end-to-end strategy to train the SQNet. In each training epoch, three TE images were stacked into a three channel image and input to the network. The SQNet was trained by minimizing the loss function in Eq. (3) with the network optimizer Adam. The learning rate was 0.001 and network parameters were initialized using the kaiming distribution. The programming language was PyTorch and the GPU was an NVIDIA A10 with 24 GB of graphics memory.

## III. RESULT

Results were validated on cardiac images acquired from a healthy control group (N=17), and an AMI patient group (N=12).

### A. Evaluation Metrics

To evaluate the segmentation accuracy, we used the Dice score as the metric [44] and performed the Bland-Altman (BA) analysis [23, 45] between SQNet and NLLS [24, 46].

In BA analysis, if a one-sample t-test using the standard normal distribution yields P>0.05, suggesting that the distribution of differences between the SQNet and NLLS in both groups are similar to the standard normal distribution and the systematic bias between the two methods is small. Besides, ± 1.96 Standard Deviation (SD) can be used as a consistency bound to show the 95% confidence interval of the distributed data.

To better understand the $T_2$ map, following the 17-segment model from the American Heart Association (AHA) [47], a bullseye is plotted to show the mean $T_2$ value of each segment area (excluding the true apex) [48]. The center of the bullseye (17[th] Segment) represents the global mean.

### B. Compared Methods

The proposed SQNet was compared with three state-of-the-art (SOTA) segmentation methods including Trans-UNet [49], Swin-UNet [50], and DF-Net [51] and one SOTA $T_2$ quantification method DeepFittingNet [23]. Trans-UNet integrated transformer self-attention into UNet for enhanced pixel associations [52]. Swin-UNet employed a Swin transformer for handling long-range dependencies in large medical images [53]. DF-Net specialized in cardiac short-axis MRI image segmentation and won outstanding segmentation performance on the MICCAI Automatic Cardiac Diagnosis Challenge (ACDC) [54]. DeepFittingNet estimated the $T_2$ value at each pixel with a 1D neural network that is composed of a recurrent neural network and a fully connected neural network [23].

### C. Segmentation Results

In the healthy volunteer image (Fig. 4(a)), over-smooth (arrow in the 3[rd] column of Fig. 4(a)) was introduced by Trans-UNet, and over rough (arrow in the 4[th] column of Fig. 4(a)) in Swin-UNet led to errors in LV and MYO. Both DF-Net and the proposed SQNet segmented closer regions to the ground-truth ones.

In the AMI patient image (Fig. 4(b)), the details of MYO were missing (arrow in the 3[rd] and 5[th] column of Fig. 4(b)) in

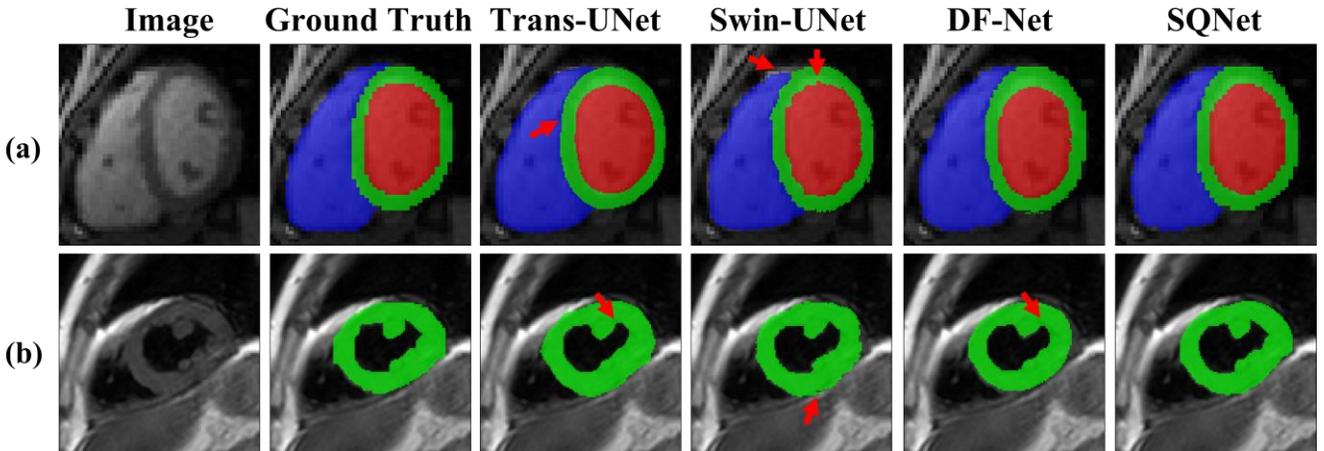

**Fig. 4.** Representative segmention results between the proposed and compred state-of-the-art methods. (a) a bright blood image of a healthy subject; (b) a black blood image of an AMI subject. Red, green and blue color denotes the region of RV, MYO and LV, respectively. Because there is no value in the LV and RV of black blood AMI, only the MYO area is drawn in (b).

TABLE II
SEGMENTATION DICE SCORE (%) FOR DIFFERENT METHODS

| Areas | Methods | Trans-UNet | Swin-UNet | DF-Net | SQNet |
|---|---|---|---|---|---|
| Left Ventricle | Healthy control | 92.6 ± 4.4 | 92.2 ± 10.7 | 91.1 ± 5.3 | **92.7** ± 4.5 |
| | AMI | 88.0 ± 6.2 | 86.8 ± 10.7 | 90.0 ± 5.9 | **91.1** ± 3.8 |
| Myocardium | Healthy control | 79.1 ± 9.0 | 74.9 ± 13.7 | 76.1 ± 8.5 | **81.9** ± 7.1 |
| | AMI | 84.8 ± 4.5 | 83.3 ± 4.9 | 86.3 ± 3.2 | **86.8** ± 3.6 |
| Right Ventricle | Health | 91.3 ± 5.9 | 89.4 ± 5.5 | 91.9 ± 3.8 | **93.3** ± 4.7 |
| | AMI | 87.1 ± 5.9 | 82.7 ± 5.7 | 87.4 ± 7.1 | **90.4** ± 5.2 |
| Average | Healthy control | 87.7 ± 8.9 | 84.8 ± 12.9 | 86.3 ± 9.6 | **89.3** ± 7.6 |
| | AMI | 86.6 ± 5.6 | 84.3 ± 7.7 | 87.9 ± 5.7 | **89.2** ± 4.6 |

Note: The health group ($N$=17), and the AMI group ($N$=12). The table presents the average and standard deviation of the Dice scores for each method, with bold indicating the highest Dice score.

Trans-UNet and DF-Net, and over rough (arrow in the 4$^{th}$ column of Fig. 4(b)) was also emerged in Swin-UNet. However, these phenomena were not observed in SQNet, indicating much better segmentation in AMI patients.

The Dice score in Table II showed that, on healthy volunteer images, SQNet performed better than Trans-UNet, Swin-UNet, and DF-Net by 2.1%, 5.0%, and 3.0%, respectively. On AMI patient images, SQNet obtained a higher Dice than Trans-UNet, Swin-UNet, and DF-Net by 2.6%, 4.9%, and 1.3%, respectively.

Overall, SQNet achieved a higher average of Dice and a smaller standard deviation of Dice than other methods, indicating more accurate segmentation and lower variation in all cases.

*D. Quantification Results*

For the analysis of the T$_2$ map in the MYO region, both BA analysis and Pearson correlation coefficient were adopted to demonstrate the agreement and correlation between the proposed SQNet and NLLS, on both healthy volunteers and AMI datasets.

In the healthy volunteers (Fig. 5(a)), BA analysis revealed excellent agreement, with a mean bias of 1.4 ms while the upper and lower 95% limits of agreement of 5.6 ms and -2.8 ms, respectively. The coverage was 94%. The Pearson correlation coefficient was 0.84 ( $\geq$ 0.8), indicating a strong linear correlation in myocardial T$_2$ mapping.

In the AMI dataset (Fig. 5(b)), the mean bias was 1.8 ms, with the upper and lower 95% limits of agreement of 5.2 ms and -1.6 ms, respectively. The coverage was 94%. The Pearson correlation coefficient was 0.93 ($\geq 0.8$), indicating a stronger linear correlation in myocardial $T_2$ mapping.

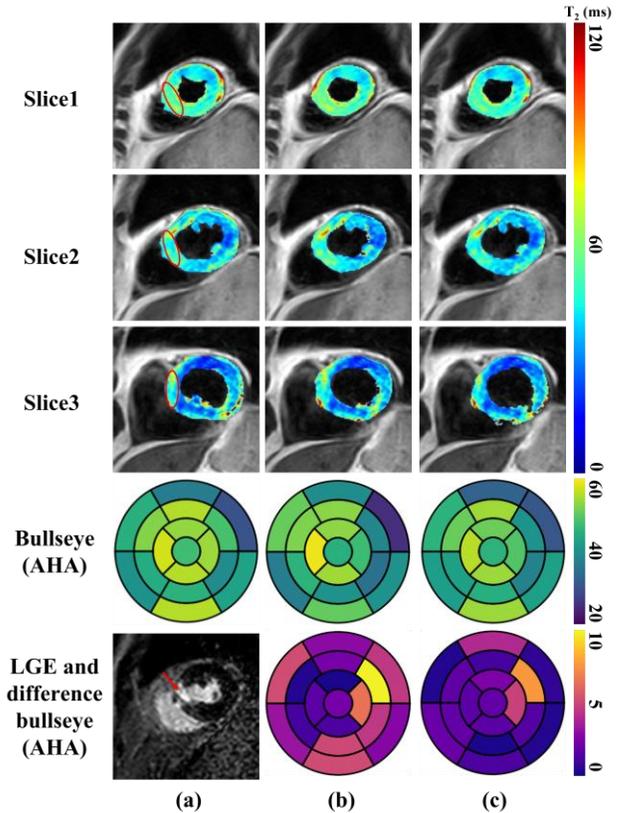

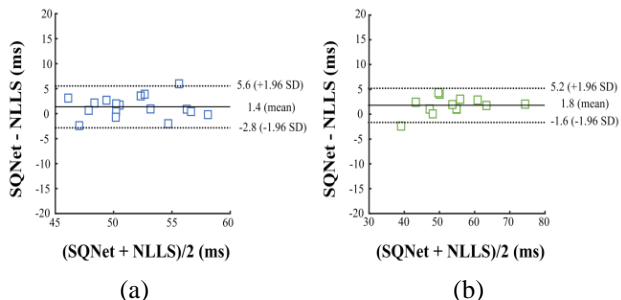

**Fig. 5.** The BA plots of T$_2$ mapping of the MYO region between the SQNet and the NLLS in (a) health group ($N$=17) and (b) AMI group ($N$=12). The difference between the two measurements is plotted on the y-axis and the corresponding mean is plotted on the x-axis. The mean of difference is denoted as solid lines and limits of agreement is denoted as dashed lines. SD means the standard deviation.

**Fig. 6.** A representative result of simultaneous quantification and segmentation on an AMI subject. (a) denotes the ground-truth T$_2$ mapping and segmentation, (b) denotes the T$_2$ mapping result of DeepFittingNet with the segmentation of DF-Net, and (c) denotes the T$_2$ mapping and segmentation results of SQNet. The center of the bullseye represents the global mean. Red circles and arrows indicate edematous areas of high signal intensity on the T$_2$ mapping and LGE images, respectively. Difference bullyseye is calculated by doing residual between two methods (b,c) and the first column (a) reference bullseye.

A representative quantification result of an AMI patient was shown in Fig. 6. The infarct edema region was located (arrow in the 5th row of Fig. 6(a)) through the Late Gadolinium Enhancement (LGE) scanning. Correspondingly, the $T_2$ value at this region was relatively higher (circles in the 1st to 3rd row of Fig. 6(a)) than other regions of the myocardium. Compared with the SOTA result (the 5th row of Fig. 6(b)), which was obtained with the SOTA combination of the segmentation method (DF-Net) and the quantification method (DeepFittingNet), the proposed SQNet achieved much lower quantification error (the 5th row of Fig. 6(c)) in the bullseye plot of $T_2$ mapping.

## IV. DISCUSSIONS

### A. Reader Study

A reader study was conducted from a diagnostic perspective. The clinical evaluation is divided into two groups. The first group included four radiologists with 10/15/23/31 years of image diagonosis experience and three cardiac specialized radiologists with 8/8/20 years of image diagonosis experience. The second group included two radiologists with 15/21 years of work experience and one cardiac specialized radiologist with 8 years of work experience. Readers were independently and blindly to each method and the order of methods was random.

TABLE III
5-POINT LIKER SCALE SCORING CRITERIA FOR THE READER STUDY

| Score | Segmentation | $T_2$ mapping |
|---|---|---|
| 5 | Excellent | Excellent |
| 4 | Good | Good |
| 3 | Consistent | Adequate |
| 2 | Inconsistent | Poor |
| 1 | Deficient | Non-diagnostic |

TABLE IV
THE SCORE OF THE READER STUDY [MEAN ± STD]

| Method | Segmentation (Healthy control/AMI) | $T_2$ Mapping (Healthy control/AMI) |
|---|---|---|
| DF-Net | 4.50 ± 0.73/ 4.44 ± 0.72 | Not available |
| DeepFittingNet | Not available | 3.59 ± 0.97/ 4.37 ± 0.68 |
| Ablation | 1.79 ± 1.08/ 3.49 ± 1.21 | 1.61 ± 1.25/ 1.47 ± 1.05 |
| SQNet | **4.60 ± 0.55/ 4.58 ± 0.56** | **4.32 ± 0.59/ 4.42 ± 0.62** |

Note: The means and standard deviations were computed over all images. Numbers before and after "/ "denotes the subjective scores on healthy volunteers and AMI patients, respectively. Highest scores are bold faced. Wilcoxon signed-rank test P-Value between SQNet of Segmentation and DF-Net is 0.033, between SQNet of $T_2$ mapping and DeepFittingNet is 0.000.

The images used for the reader study included 17 healthy subjects and 12 AMI subjects. Each subject included 3 short-axis slices. We used confidence as the evaluation criteria for both segmentation and $T_2$ mapping. The score was ranged from 0 to 5 (Table III). The Wilcoxon signed-rank test was employed to analyze scores between SQNet and other methods.

Ready study (Table IV) showed that the SQNet reached a segmentation score of 4.60 (a $T_2$ mapping score of 4.32) for healthy controls and a segmentation score of 4.58 (a $T_2$ mapping score of 4.42) for AMI patients. These scores lied in the range of excellence (4.0~5.0). Besides, these scores were much higher than those of the ablation of SQNet (removed either the segmentation or quantification branch of the proposed SQNet), implying the necessity of simultaneous dual-task learning.

Besides, SQNet also obtained higher scores than the SOTA segmentation method (DF-Net) and quantification method (DeepFittingNet). The Wilcoxon signed-rank test showed that SQNet had significant advantages ($P<0.05$) over DF-Net and DeepFittingNet in segmentation and $T_2$ mapping, respectively.

### B. Ablation Study

Table V summarized the segmentation results under the ablation study. When removing the $T_2$ mapping module, the average segmentation dice score of SQNet was reduced by 0.3% and its standard deviation was increased for the healthy group but reduced for the AMI group. This observation implied that $T_2$ mapping can improve the segmentation but the improvement was slight.

Fig. 7 showed the $T_2$ mapping results under the ablation study. When removing the segmentation module, the mean bias was increased from 1.4 ms (Fig. 5(a)) to 2.1 ms (Fig. 7(a)) in the healthy volunteers, and from 1.8 ms (Fig. 5(b)) to 2.9 ms (Fig. 7(b)) in the AMI datasets. This observation implied that segmentation can highly improve the $T_2$ mapping.

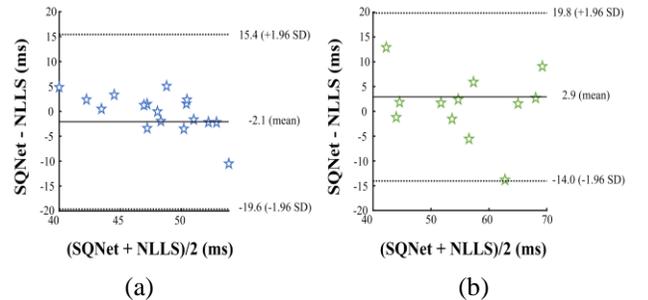

**Fig. 7.** The BA plots between the SQNet and the NLLS fit the $T_2$ mapping of the MYO region in (a) health group ($N$=17) and (b) AMI group ($N$=12). The difference between the two measurements is plotted on the y-axis and the corresponding mean is plotted on the x-axis. The mean of difference is denoted as solid lines and limits of agreement is denoted as dashed lines. SD means standard deviation.

### C. Noise Adaptation Experiment

Noise adaptation was tested by adding Gaussian noise into images. The noise was categorized into three levels: low (STD=0.01), medium (STD=0.03), and high (STD=0.05).

TABLE V
ABLATION STUDY: SEGMENTATION DICE SCORE (%) FOR SEGMENTATION AND T$_2$ MAPPING TASKS

| Models | Segmentation (CNN) | T$_2$ mapping (Transformer) | Average Dice score | |
|---|---|---|---|---|
| | | | Healthy control | AMI |
| #1 | √ | × | 89.0 ± 7.0 | 88.9 ± 5.6 |
| #2 | √ | √ | **89.3** ± 7.6 | **89.2** ± 4.6 |

Note: Cardiac images were acquired from a healthy control group (N=17), and the AMI patient group (N=12). The table presents the average and standard deviation of the Dice scores for each method, with bold indicating the highest Dice score.

TABLE VI
SEGMENTATION DICE SCORE (%) UNDER DIFFERENT NOISE LEVELS BETWEEN OUR METHOD AND OTHER STATE-OF-THE-ART METHODS

| STD | Methods | Trans-UNet | Swin-UNet | DF-Net | SQNet |
|---|---|---|---|---|---|
| Low | Healthy control | 87.7 ± 8.8 | 83.2 ± 12.9 | 87.9 ± 7.2 | **89.2** ± 6.8 |
| | AMI | 87.0 ± 6.2 | 82.1 ± 8.8 | 86.7 ± 6.4 | **87.9** ± 5.2 |
| Mid | Healthy control | 88.2 ± 8.2 | 84.0 ± 12.4 | 87.9 ± 8.2 | **89.1** ± 6.5 |
| | AMI | 87.7 ± 5.2 | 81.7 ± 8.2 | 88.3 ± 6.0 | **89.3** ± 4.0 |
| High | Healthy control | 88.2 ± 8.4 | 83.0 ± 15.8 | 86.2 ± 11.0 | **89.2** ± 7.2 |
| | AMI | 87.0 ± 6.2 | 78.9 ± 14.0 | 88.3 ± 5.7 | **88.5** ± 4.8 |

Note: The healthy control group (N=17), and the AMI group (N=12). The table presents the average and standard deviation of the Dice scores for each method, with bold indicating the highest Dice score. Low gauss noise std is 0.01, mid Gaussian noise std is 0.03, and high gauss noise std is 0.05.

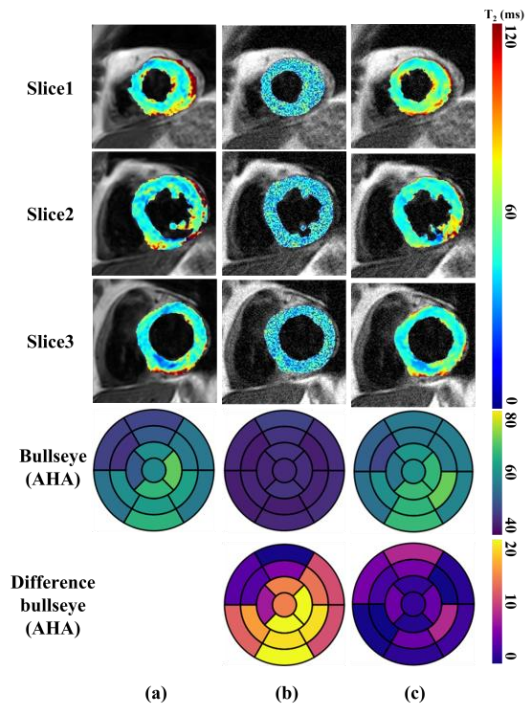

**Fig. 8.** Simultaneous segmentation and quantification of a noise adaptation experiment. (a) denotes the T$_2$ mapping result of NLLS with the ground truth segmentation, (b) denotes the high noise level segmentation result of DF-Net with a high noise level fitting of DeepFittingNet, (c) denotes the SQNet simultaneous segmentation and quantification results under high noise level. Difference bullyseye is calculated by doing residual between two methods (b,c) and the first column (a) reference bullseye.

The network label of segmentation and T$_2$ mapping were obtained from the noise-free images. The network input was the noise-add images. All methods were re-trained under different noise levels.

Table VI shows that SQNet still had the highest segmentation accuracy under three noise levels on datasets. Fig. 8 showed that SQNet can maintain better analysis results at high noise levels and has the smallest error (the 5$^{th}$ row of Fig. 8(c)) on the bullseye plot. Thus, the SQNet allowed strong resistance to noise and robustness.

## V. CONCLUSION

In this work, we proposed the SQNet, a deep learning-based dual-task network tailored for myocardium MRI segmentation and T$_2$ quantification. Objective and reader-based subjective evaluations on the healthy controls and AMI patients data demonstrate that both tasks can be successfully achieved and the accuracy of T$_2$ quantification can be improved.